\documentclass[letterpaper,twocolumn,10pt]{article}
\usepackage{usenix2019_v3}

\usepackage{tikz}
\usepackage{amsmath}

\usepackage{filecontents}

\usepackage{array,graphicx,multirow}
\graphicspath{{./Figures/}}

\usepackage{comment}

\usepackage{xspace}
\newcommand{\vs}{vswitch\xspace}
\newcommand{\vswvm}{vswitch VM\xspace}

\newcommand{\sriov}{SR-IOV\xspace}

\newcommand{\vm}{VM\xspace}

\newcommand{\host}{Host\xspace}

\begin{document}

\date{}
\title{\Large \bf Making The Case For Multi-Tenant Virtual Switches}
\title{\Large \bf Making The Case For Tenant-Specific Virtual Switches}
\title{\Large \bf Towards Fine-Grained Billing For Cloud Networking}

\author{
} 

\author{
{\rm Kashyap Thimmaraju}\\
Security in Telecommunications\\
Technische Universit{\"a}t Berlin
\and
{\rm Stefan Schmid}\\
Faculty of Computer Science\\
University of Vienna
} 

\maketitle

\begin{abstract}
We revisit multi-tenant network virtualization in data centers,
and make the case for {\em tenant-specific} virtual switches.
In particular, tenant-specific virtual switches
allow cloud providers to extend fine-grained billing (known, e.g.,
from serverless architectures) to the {\em network},
accounting not only for IO, but also CPU or energy. 
We sketch an architecture and present economical motivation
and recent technological enablers.
We also find that virtual switches today do not offer sufficient multi-tenancy 
and can
introduce artificial performance bottlenecks, e.g., in load
balancers. 
We conclude by discussing additional use cases for tentant-specific switches.
\end{abstract}

\section{Introduction}
\label{sec:intro}

Network virtualization in data center networks has come a long way since the
initial days of using Linux bridge to interconnect virtual network interfaces of
different (tenant) virtual machines (\vm{}s)~\cite{pfaff2009extending}. The
state-of-the-art uses sophisticated software switches called virtual
switches~\cite{pfaff2015design} (\vs{}) that enforce isolation between the different \vm{}s on the
server as well as across servers using e.g., a tunnelling protocol like VXLAN.

This general design of co-locating the \vs{} with the \host{},
has been adapted and tailored to the
specific needs, workloads and goals of various cloud providers. For example,
Azure's VFP~\cite{firestone2017vfp} and AccelNet~\cite{accelnet} uses a hybrid solution of a
software switch accelerated by a hardware network card (NIC); Google Cloud
Platform~\cite{andromeda-2018}, uses a software-only solution for their virtual
switch, with additional network services either offloaded to Hoverboards or
Coprocessors running in dedicated servers or co-located \vm{}s respectively.

\begin{figure}[t!]
    \centering
    \includegraphics[scale=0.150]{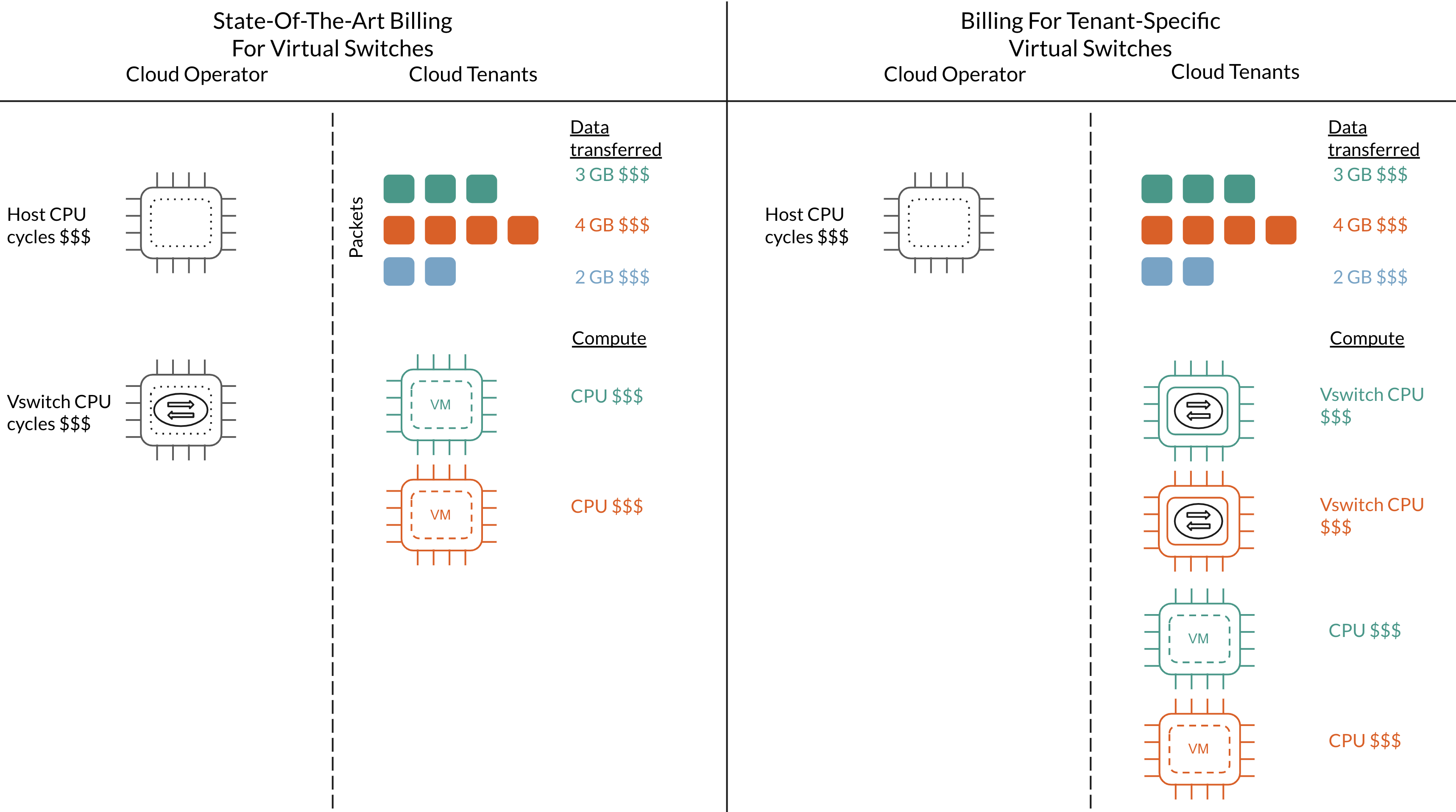}
	\caption{Comparison of the state-of-the-art billing for virtual networking
and tenant-specific \vs{}es. In the latter, the cloud operator can charge the
tenant for CPU cycles consumed by the \vs{} and the tenant only pays for the
\vs{} she uses, which was previously not possible.}
    \label{fig:fine-grain-pricing}
\end{figure}

This fate-sharing model scales linearly with the number of servers and is
somewhat manageable. However, it comes with restrictions, performance
limitations and security weaknesses, e.g.:

\begin{itemize}
    \item If all tenants share the underlying \vs{}, they are
restricted to the functionality of that particular switch: they
have to wait for an upgrade/update from the cloud provider/software switch vendor. 
Furthermore,
because the \vs{} is co-located with the host, the \emph{cloud operator
needs to bear the cost of operating the \vs{}}~\cite{ec2-pricing, azure-netpricing, gcp-netpricing}.

    \item High-performance software network functions that outperform the
\vs{} are limited to the performance of the \vs{}, e.g.,
software load-balancers can process packets are rates higher than the virtual
switch, however, they are limited to the performance of the virtual
switch~\cite{avi-networks-packetpushers-podcast}. Software load-balancer vendors
(e.g., Avi Networks) have suggested a workaround for this by using dedicated
servers for tenant-specific load balancing. 

    \item From a security perspective co-locating the \vs{} with the
Host compounded by its monolithic code and configuration is vulnerable to
compromise~\cite{ovs-sosr,vamp,181358,sv3,csikor2019tuple}.
\end{itemize}

In this paper, we make the case for deploying
\emph{tenant-specific} \vs{}es in the cloud,
which can overcome the aforementioned issues.
Our approach is enabled by recent developments~\cite{mts} showing 
promising tradeoffs between security, performance and
resources of multi-tenant \vs{}es (Section~\ref{sec:discussion}).

In particular, we argue that tenant-specific virtual switches enable cloud
providers to amortize the costs of virtual networking by fine-grained accounting
and pricing (as illustrated in Figure~\ref{fig:fine-grain-pricing}), e.g., by
accounting for the CPU cycles used by tenant-specific \vs{}es; also IO and
energy can be factored in.  Such mechanisms are particularly well-suited for the
serverless architecture~\cite{ucb-serverless}, where billing is naturally
fine-grained, i.e., by the second rather than by the hour or usage-based. The
main challenge we believe cloud providers currently face in achieving such
fine-grained pricing is the lack of a method to perform such accounting.

For example, the cloud provider Snowflake recently mentioned the
need to rethink and redesign their cloud systems to perform
fine-grained billing~\cite{fine-grain-pricing-motivation}.
Furthermore, Singhvi et al.~\cite{singhvi2017granular} describe
the lack of fine-graining billing for the networking in Serverless
architectures such as AWS Lambda. Indeed, the pricing for networking
for AWS Lambda merely adopts the hourly billing rates~\cite{aws-lambda-net}.
In NFV, the programs are naturally network and CPU intensive hence,
SNF~\cite{singhvi2019snf} proposed by Singhvi et al.,
uses a per-packet granularity for billing serverless NFV
applications. However, in SNF, the cloud operator still pays
for CPU cycles used by the \emph{single} \vs{} on the server.

The remainder of this paper is organized as follows. In Section~\ref{sec:case}
we make our case for tenant-specific virtual networking by first describing the
concept of tenant-specific virtual networking and then estimating the cost
savings of such an architecture for the cloud operator and tenants. Next, in
Section~\ref{sec:discussion} we discuss the technical aspects of our proposal.
Finally, we highlight additional use cases in Section~\ref{sec:possibilities}.


\section{The Case for Tenant-Specific Switches}
\label{sec:case}

\begin{figure}[t!]
    \centering
    \includegraphics[width=0.45\textwidth]{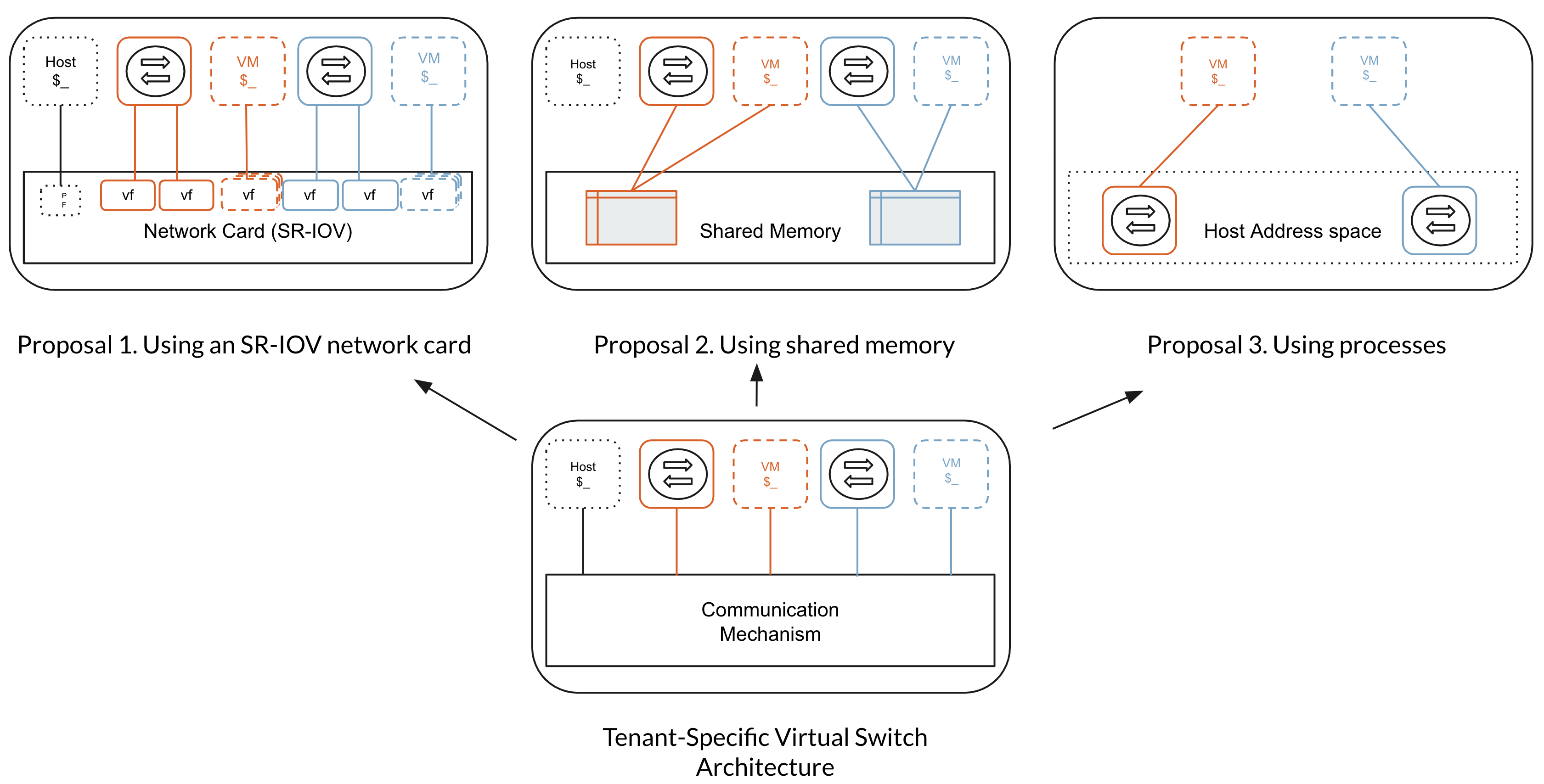}
	\caption{Conceptual illustration of tenant-specific \vs{}es (bottom) along with three
different approaches (top).}
    \label{fig:concept}
\end{figure}

The concept of tenant-specific virtual networking can be discerned via the
illustration in Figure~\ref{fig:concept}. Virtual switches dedicated to tenants
run as compartments (e.g., processes, \vm{}s, containers, etc.). Through a
communication medium (e.g., shared memory, PCIe bus) and mechanism,
the \vs{} and
its tenants can exchange network packets. Via a logically centralized
controller, the cloud provider controls (and monitors) the communication channel
between the \vs{} compartments and their tenants, as tenant isolation
is critical. The cloud provider also configures the tenant-specific virtual
switch. Resources, in particular, CPU cores for the \vs{}es can be
allocated in several ways: all \vs{}es are pinned to a dedicated core;
each \vs{} gets a dedicated core; the \vs{} and the
respective tenant \vm{}s share the same core; or a 
combination of \vs{}es
and cores.

The ability to (flexibly) allocate cores to tenant-specific \vs{}es, now enables
the cloud provider to precisely measure, e.g., the number of CPU cycles, cache
lines, pages in memory or energy consumed whenever a \vs{} compartment is
scheduled as illustrated in Figure~\ref{fig:fine-grain-pricing} where CPU cycles
from the tenant-specific \vs{} are billed.  In this way, the resources consumed
by the \vs{} compartment can be amortized by charging the customer accordingly.
Note that, this also benefits the tenant as she does not have to pay for network
communication processing she does not use, hence, a potential win-win for the
operator and tenants.

\subsection{Economic Motivation}
\label{sec:economics}

To illustrate the potential economic benefits of a tenant-specific \vs{} design,
we discuss a back-on-the-envelope calculation for the cloud operator and cloud
tenants.

\subsubsection{Cloud Operator}
\label{sec:eco-cloud}
We compare the state-of-the-art with three options:
\begin{itemize}
\item\textbf{Option 1:} one dedicated core for all tenant \vs{}es.
\vspace{-.8em}
\item\textbf{Option 2:} the \vs{} shares the resp. tenant's cores.
\vspace{-.8em}
\item\textbf{Option 3:} one dedicated core for each tenant's \vs{}.
\end{itemize}

We make the following simplifying 
assumptions as we do not have access to the operational costs
and pricing models of cloud providers. Nonetheless, we use the publicly
available prices from Amazon EC2~\cite{ec2-pricing} for the pricing of \vm{}s.

\noindent\textbf{Without tenant-specific switches.}
A compute instance (\vm{}) costs 1 cent/hour (or .01 cent/second) and we consider 
100 servers, each with 
12 cores. 
Each server can host 10 \vm{}s (2 cores for the Host)
and 
hosts a maximum of 5 tenants (i.e., a max of 2 tenant \vm{}s per
server).
Thus, in 24 hours (1 day), the total
income can amount to 1*24*10*100 = 24,000 cents/day = 
87,600 dollars/year.
The money spent in 24 hours 
is 2*24*100 = 4800 cents/day = 
17,520 dollars/year.
The total revenue is hence
the difference, 70,080 dollars/year.


\noindent\textbf{Tenant-specific switches: Option 1.}
In this case we have the following.
Each server continues to host 10 \vm{}s, however,
only 1 core is allocated to the Host and the
other core is shared by all the tenant-specific \vs{}es.
We assume the vswitch \vm{}s are used only 50\% of the time compared to the workloads.
Thus, in 1 day, the total income from the \vs{} \vm{}s can amount to
1*12*1*100 = 1,200 cents/day = 4,380 dollars/year.
The money spent reduces by half compared to the previous calculation
as the Host has only 1 core now, which is now 8,760 dollars/year.
The net revenue is hence, 4,380+87,600-8,760 = 83,220 dollars/year.
This is a net revenue of 18.75\% more compared to co-locating the virtual
switch with the Host and bearing the cost for 2 cores.

\noindent\textbf{Tenant-specific switches: Option 2.}
In this scenario, the \vs{} \vm{} does not need any extra core
as it is shared with the tenant's \vm{}s. This allows us to take 1 core
from the Host and run an extra tenant with a single \vm{} on each server.
Hence, we have the following income,
1*24*11*100 = 26,400 cents/day = 96,360 dollars/year.
The money spent is similar to Option 1, which results in a
net revenue of 96,360-8,760 = 87,600 dollars/year which is
25\% more to the state-of-the-art.

\noindent\textbf{Tenant-specific switches: Option 3.}
Here, the situation is a little more complicated as each virtual
switch \vm{} gets a dedicated core. This results in tenants being displaced
which requires purchasing new servers to host the displaced tenants.
Based on that, we have the following.
Similar to the previous two options, with the Host being allocated 1 core,
we have an extra core.
Therefore, each server can now host 3 tenants, each tenant consumes
3 cores: 2 for the workload and 1 for the vswitch \vm{}.
The remaining 2 cores can be allocated for a tenant with a single \vm{}.
Hence, we have 4 tenants per server: 7 (workload) \vm{}s and 4 vswitch \vm{}s.
This generates 1*24*7*100 = 16,800 cents/day = 61,320 dollars/year
from the tenant \vm{}s.
In addition, the \vs{} \vm{}s (running at 50\%) generate
1*12*4*100 = 4,800 cents/day = 17,520 dollars/year.
The total money earned from the 100 servers = 78,840 dollars/year.
However, as mentioned earlier, by dedicating cores for the vswitch \vm{}s,
we have 1.5 tenants displaced per server, or 1.5*100 = 150 tenants
displaced in total. Since, each server can host 3.5 tenants, we will
need 43 new servers, where each server costs 2,000 dollars.
From the 43 new servers, we have the following earned.
1*24*7*43 + 1*12*4*43 = 7,224+2,064 = 9,288 cents/day = 33,901.2 dollars/year.
The total income is 112,741.2 dollars/year.
The money spent from the 143 servers is 12,526.8 dollars/year.
The net revenue is hence 100,214.4 dollars/year without considering
the captial/sunk cost of the 43 servers which is 43*2,000 = 86,000 dollars.
For simplicity we do not consider a return on investment calculation here.
This is 43\% more than the state-of-the-art.

\noindent\textbf{Key takeaway.} Based on our back of the envelope
calculations we find that dedicating tenant-specific \vs{} \vm{}s/compartments
with billable CPU resources introduces potential savings for the cloud provider
from 10\% upto 40\%.  Usage-based billing or pricing per millisecond could offer
more savings.

\subsubsection{Cloud Tenant}
\label{sec:eco-tenant}
As we just saw the potential savings for the cloud operator, we now cast light
on the savings for cloud tenants based on the three options described for the
cloud operator. In particular, we assume the 5 tenants co-located per server
use their \vs{} \vm{} as follows: 1\%, 2\%, 2\%, 5\%, 10\% and 30\%.
This adds up to the 50\% usage as stated previously for the cloud operator.

\noindent\textbf{Tenant-specific switches: Option 1, 2 and 3.}
In Options 1 and 3, the savings per tenant are directly proportional to her time usage.
For example, the tenant who consumes 10\% pays two thirds less than the tenant who
consumes 30\% of the CPU for the \vs{} \vm{}.
In Option 2, savings manifest if billed by CPU cycles used rather than time slices.

\subsection{Technological Enablers}
\label{sec:enablers}

Having cast light on the economic benefits of a tenant-specific \vs{}, we
now point out key technologies currently available in the market that enable
such a system.

Compartments can trivially be realized via virtual machines, containers or
processes as operating systems and hardware already support these primitives
with security and performance guarantees. The main challenge lies in connecting
the tenant-specific \vs{} compartment with the tenant \vm{}s and then
enforcing isolation between the different tenant \vs{} and \vm{}s.

Broadly speaking, there are two potential solutions: software or hardware.
A software based approach has been proposed by Jin et al.~\cite{181358}, as well as
Stecklina~\cite{sv3}. The former design places the \vs{} in a \vm{}
(compartment) and then plumbs together the \vs{} with all the tenants
using shared memory. This requires modification to the hypervisor/virtual
machine monitor as well as fine tuning, e.g., polling frequency, buffer sizes,
message lengths, etc.
Furthermore, polling is expensive as the CPU is constantly being used even when
there are no packets. The latter design
describes an isolation technique that leverages
process isolation for tenant specific \vs{}es. Again the communication
between the \vs{} and the respective tenant \vm{} is stitched together via
changes to QEMU/KVM. Nonetheless, the latter approach can allow \vs{}
processes to be pinned to dedicated cores and hence billed appropriately.
Other possibilities include a language-centric apporach, e.g., using RUST.

A hardware based approach recently proposed by Thimmaraju et al.~\cite{mts} uses
off-the-shelf hardware and software for network virtualization. In particular,
the design proposes Single Root IO Virtualization (\sriov{})~\cite{mts} as most
CPUs and NICs used in the cloud already offer these features. Isolation is
enforced at the NIC via a simple layer 2 Ethernet switch using VLAN tags.
Virtual functions (VFs) on the NIC can be allocated to the \vs{} \vm{} and
tenant \vm{}s as desired. The \vs{} and the tenants are then appropriately
configured for connectivity and isolation.

\section{Technical Aspects and Realization}
\label{sec:discussion}
We now discuss the benefits and challenges we foresee in realizing and operating
a multi-tenant \vs{}es in the cloud. To aid our discussion, we use
MTS~\cite{mts} as a case study.

\begin{figure*}[t!]
    \centering
    \includegraphics[scale=.5]{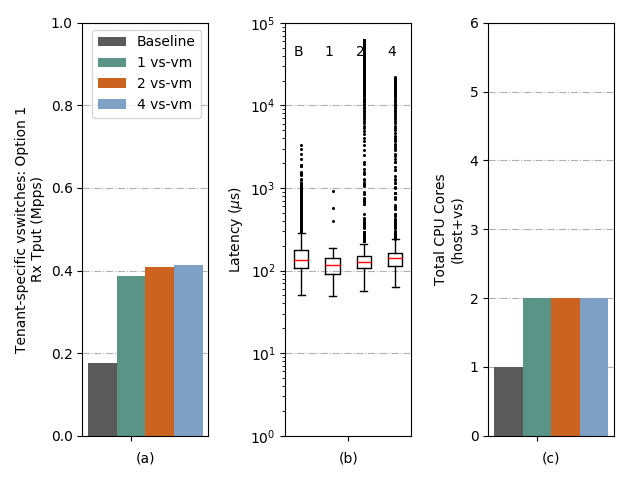}
    \includegraphics[scale=.5]{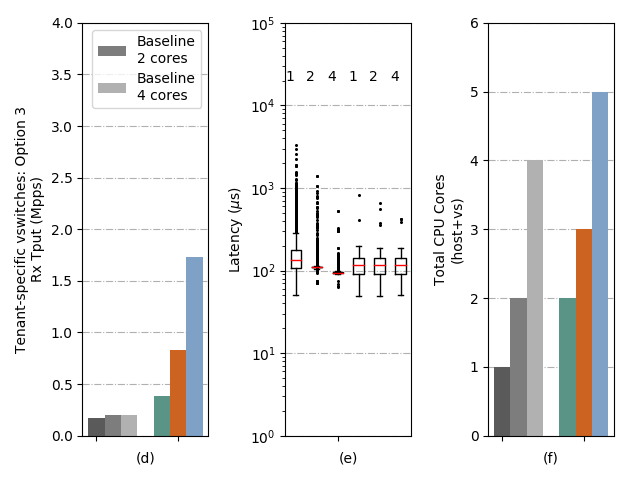}
    \includegraphics[trim=0.0cm 0.0cm 5.0cm 0.0cm,clip=true,scale=.5]{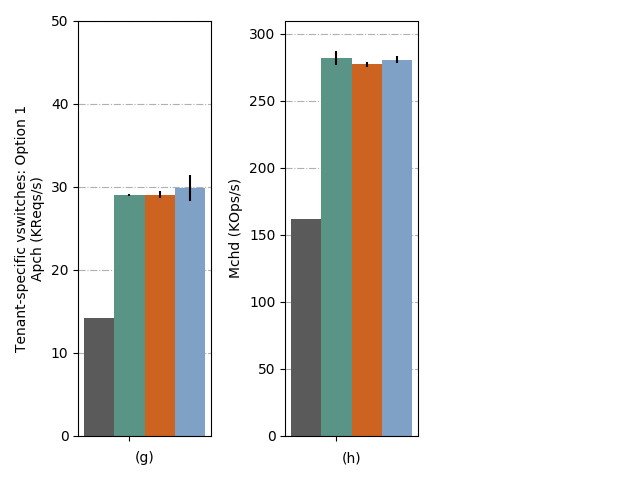}
    \includegraphics[trim=0.0cm 0.0cm 5.0cm 0.0cm,clip=true,scale=.5]{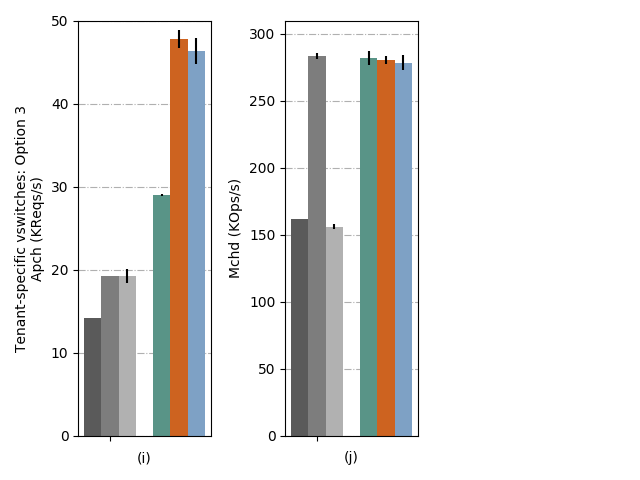}
    \includegraphics[trim=0.0cm 0.0cm 5.0cm 0.0cm,clip=true,scale=.5]{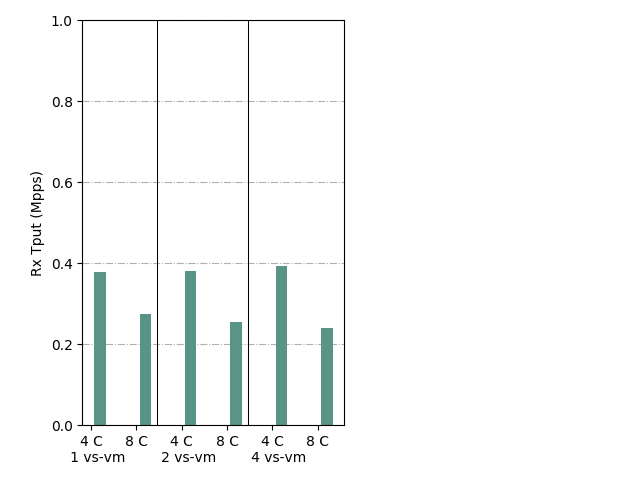}
    \caption{Performance evaluation of a tenant-specific \vs{} architecture
    using \sriov{}.
    The top row depicts the packet processing throughput, latency
    and CPU cores for tenant-specific \vs{}es Options 1 and 3 for traffic
	in the canonical p2v scenario. The bottom row depicts the throughput of
    Apache and Memcached for Options 1 and 3 also in the p2v scenario. The bottom
    right figure shows the packet processing throughput of running the \vs{}
    in containers in \vm{}s.}
    \label{fig:mts-perf}
\end{figure*}

\noindent\textbf{Security, Performance \& Resources}
First is the point of the tradeoffs between security, performance and resources.
As we can see in Figure~\ref{fig:mts-perf}, a multi-tenant architecture using
\sriov{} improves the packet processing throughput and latency for Options 1 and
3 (recall Section~\ref{sec:economics}). Furthermore, we can see that workloads
in the tenant \vm{}s, e.g., Apache and Memcached, also receive a performance
improvement. It is obvious that Option 3 consumes more cores than Option 1 as
each \vs{} \vm{} is dedicated a core. However, the performance and economic
beneifts (Section~\ref{sec:eco-cloud}) are also proportional.  Indeed, only
Option 1 and 3 were evaluated, however, it goes to show that by running the
\vs{} in a dedicated \vm{}, we can dynamically scale the security, performance
and resources for virtual networking.

\noindent\textbf{Configuration}
Virtual networking configuration and management is highly complex even if it has
been wrapped around a programmatic framework. A misconfiguration in one virtual
switch could end up leaking traffic and violating network isolation guarantees
and policies~\cite{mts}. One solution to reduce the complexity is to move away
from a monolithic configuration model to a tenant-specific model. Tenant
specific network configuration is restricted to the tenant’s \vs{}. The tradeoff
here lies in duplicating the necessary configuration across the tenant-specific
\vs{}es which costs memory. Controllers would need to be modified to operate on
tenant-specific \vs{}es in addition to configuring the communication medium
(e.g., \sriov{} NIC), as well as flexibly allocate resources to the \vs{}
compartments/\vm{}s.

\noindent\textbf{Management}
With respect to management, indeed introducing tenant-specific \vs{}es will
increase the number of \vs{}es that have to be managed by the controller.
Furthermore, this will also introduce more network traffic along the links
between the switches and controller.  From a security perspective, this would
also mean managing more keys and certificates, e.g., for TLS sessions between
the switch and controller.

\noindent\textbf{Routing}
Indeed, the way switching and routing operate will have to be adapted as tenant
\vs{}es and their workloads (\vm{}s) will need to be reachable via the server
and network fabric. In the context of MTS, this is achieved by identifying each
tenant \vs{} by its MAC address and the workloads by its IP address. This is
currently not the way, e.g., NSX-T~\cite{nsx-t} and OVN~\cite{ovn}, are
designed. Furthermore, if the servers are not aware they are on the same
Ethernet network, then classification and reachability based on tunnels IDs
would be necessary.  With programmable NICs~\cite{nica-atc19} and
P4~\cite{bosshart2014p4}, this issue could be overcome.

\noindent\textbf{Scaling}
Being able to scale the number of \vs{}es is a high priority for cloud
operators. To that end a combination of \vm{}s and containers can be used.  In
particular, we evaluated the packet processing throughput of running OvS in
containers in \vm{}s and compared that with the measurements shown in
Figure\ref{fig:mts-perf}.  We found the overhead introduced by the container
mechanisms to be less than 2\%, hence, we do not show the plot here.
Instead, in the bottom right of Figure~\ref{fig:mts-perf} we show the
packet processing throughput for 16 tenants using 4 and 8 \vs{} containers
spread across 1, 2 and 4 VMs using Option 1. We can see that doubling the
\vs{} containers reduces the throughput by nearly 25\%. This is due to an
increased number of ports per \vm{} and pushing packets across memory intense
{\tt sk\_buff} buffers coupled with just a single core shared by all the
\vswvm{}s.  Hence, Option 2 or 3 would be able to offer better performance.

\noindent\textbf{Network Virtualization Interference}
By scheduling \vs{}es in dedicated \vm{}s, we can reduce interference
problems that arise by having a single \vs{} shared by multiple
tenants. However, interference can now surface at the communication medium,
e.g., NIC level. This can be circumvented by used dedicated resources and
rate-limiting mechanisms. For example, \sriov{} NICs have dedicated registers for
the VFs and also implement rate-limiters for VFs~\cite{mellanox-sriov-security}.
The PCIe bus could also be a source of interference or even packet loss,
hence, using more lanes or newer versions~\cite{mellanox-pci4} can alleviate this problem.

\noindent\textbf{NIC Resource Limitations}
Indeed the resources on the NIC will prove to be a limitation when using \sriov{}.
However, this is really problematic only when the degree of co-located tenants
on the same server is very high.  Microsoft azure has a limit of 60
\vm{}s/server and a total of 700 \vm{}s across their
cloud~\cite{azure-stackhub-limit}.  If we assume cloud
providers use high-end Intel xeonPhi-based servers which have 72 cores per
processor, then we are likely to hit a VF limit before the core limit: MTS
offers 21 unique tenant vswitch \vm{}s per Physical Function (port). To
reach 72, we need at least 4 PFs. Over 70 ports communicating over the PCIe bus
and NIC switch would introduce an overhead to the performance and resources of
the server.  One workaround is to limit the number of co-located tenants per
server. For example, spread them across servers and avoid the worst case VF
limitation. Note that the limitation we are talking about here is where we have
21 unique tenant vswitch \vm{}s per server. Indeed, this raises our next point
which is an increase in the number of servers to host tenants.

\noindent\textbf{Workload Displacement}
If the number of tenants per server are to be capped, then more physical servers
are necessary to host the tenants that are displaced. This would incur
additional capital costs: new servers, ports on switches, cabling, energy cooling and
power, etc. However, as mentioned in Section~\ref{sec:economics}, these costs could
be amortized by fine-grained billing of tenant-specific \vs{}es.
Furthermore, when workloads are displaced, some tenants could have to pay for
more \vs{} \vm{}s than others: if there is only 1 \vs{} \vm{} and 1
workload \vm{} on a server. Hence, placing tenants in a fair and efficient manner
across the servers would need to be devised.

\section{Additional Use Cases}
\label{sec:possibilities}

The possibility to run a \vs{} on a per-tenant basis 
has additional aspects and opportunities. 

\noindent\textbf{Bring Your Own Switch}
Using tenant-specific switches, cloud operators and the tenants do not have to wait
for a feature to be introduced into the \vs{} before they can use it.
Furthermore, a bug in one \vs{} should not impact other tenants. Hence,
being able to support tenant-specific network virtualization features by either
having tenants bring in their own \vs{} or using \vs{}es
tailored to a tenant’s needs, can make the overall system more robust and 
potentially even offer better performance.

\noindent\textbf{Snapshots And Migration}
Being able to make snapshots of the \vs{} state and configuration
enables backup and migration. For example, if there is maintenance on a
particular server or a set of servers, the cloud provider can simply take a
snapshot of the \vs{} \vm{}s and then migrate the \vm{}s to another server.
This could prove to be faster and perhaps more reliable than having a sudden
peak in traffic to the controller to update the \vs{} state and
configuration.

\noindent\textbf{Co-locate Network Functions}
Spinning up tenant-specific \vs{} \vm{}s could lead to a simplified way to
deploy tenant specific service function chains. For example, running the
different network functions including the \vs{} as containers
co-located in the \vm{} reduces the amount of traffic that has to pass through the
Host for switching. This also reduces the latency between the \vs{} and
network functions and potentially avoids any interference from other tenants.

\bibliographystyle{plain}
\bibliography{hotcloud20}

\end{document}